\documentclass[preprint,amsmath,amssymb,11pt,english,aps,showkeys,showpacs]{revtex4}
\usepackage[english]{babel}
\usepackage{graphicx}
\usepackage{graphics}
\usepackage{amsmath}
\usepackage{dcolumn}
\usepackage{amssymb}
\usepackage{bm}

\begin{document}
\title{He-McKellar-Wilkens-type effect from the Lorentz symmetry breaking effects}
\author{A. G. de Lima}
\email{andreglima@gmail.com}
\affiliation{Departamento de F\'isica e Qu\'imica, Universidade Federal do Esp\'irito Santo, Av. Fernando Ferrari, 514, Goiabeiras, 29060-900, Vit\'oria, ES, Brazil.}

\author{H. Belich} 
\email{belichjr@gmail.com}
\affiliation{Departamento de F\'isica e Qu\'imica, Universidade Federal do Esp\'irito Santo, Av. Fernando Ferrari, 514, Goiabeiras, 29060-900, Vit\'oria, ES, Brazil.}

\author{K. Bakke}
\email{kbakke@fisica.ufpb.br}
\affiliation{Departamento de F\'isica, Universidade Federal da Para\'iba, Caixa Postal 5008, 58051-970, Jo\~ao Pessoa, PB, Brazil.}

\begin{abstract}
From the effects of the Lorentz symmetry violation in the CPT-even gauge sector of Standard Model Extension, we establish a possible scenario where an analogue of the He-McKellar-Wilkens effect can stem from. Besides, we build quantum holonomies associated with the analogue of the He-McKellar-Wilkens effect and discuss a possible analogy with the holonomic quantum computation [P. Zanardi and M. Rasetti, Phys. Lett. A {\bf264}, 94 (1999)]. Finally, we investigate the dependence of the energy levels on the He-McKellar-Wilkens geometric phase induced by Lorentz symmetry breaking effects when the particle is confined to a hard-wall confining potential.
\end{abstract}


\maketitle

\section{Introduction}

It is well-known in the current literature that the Standard Model (SM) of particle physics has not gotten success of explaining the origin of electron electric dipole moment (EDM), $d_{e}$, and its experimental upper bounds \cite{revmod}. As an example, by using a polar molecule thorium monoxide (ThO), experiments measured an upper limit given by $d_{e}\leq10^{-29}\,\mathrm{e}\cdot\mathrm{cm}$ with a great confidence \cite{science}. On the other hand, studies based on the Standard Model have established an upper limit given by $d_{e}\leq10^{-38}\,\mathrm{e}\cdot\mathrm{cm}$ \cite{revmod}. Therefore, with this experimental result, it is necessary to investigate the physics beyond the Standard Model because the term associated with the electric dipole moment violates the CP symmetry. A possible way of dealing with a scenario beyond the Standard Model is the extension of the mechanism for spontaneous symmetry breaking through vector or tensor fields, which implies that the Lorentz symmetry is violated.

The research in models that present the breaking of the Lorentz symmetry has begun after the seminal work made by Kosteleck\'{y} and Samuel \cite{extra3} in the string theory, where it is shown that the Lorentz symmetry is violated through a spontaneous symmetry breaking mechanism triggered by the appearance of nonvanishing vacuum expectation values of nontrivial Lorentz tensors. Models of the Lorentz symmetry breaking are considered as effective theories whose analysis of the phenomenological aspect at low energies may provide information and impose restrictions on the fundamental theory in which they stem from. With the progress of a possible generalization of the Standard Model including the spontaneous violation of the Lorentz symmetry, thus, a general framework for testing the low-energy manifestations of CPT symmetry and the Lorentz symmetry breaking has been established in recent decades, where it is known as the Standard Model Extension (SME) \cite{colladay-kost}. Besides, in this framework, the effective Lagrangian corresponds to the usual Lagrangian of the Standard Model to which is added to the Standard Model operators a Lorentz violation tensor background. The effective Lagrangian is written as an invariant under the Lorentz transformation of coordinates in order to guarantee that the observer independence of physics. However, the physically relevant transformations are those that affect only the dynamical fields of the theory. These changes are called particle transformations, whereas the coordinate transformations (including the background tensor) are called the observer transformations. In Refs. \cite{coll-kost,baeta,bras}, these concepts are more deeply analysed. Concerning the experimental searches for the CPT/Lorentz-violation signals, the generality of the SME has provided the basis for many investigations. In the flat spacetime limit, empirical studies include the Higgs \cite{higgs} sector. The gravity sector has also been explored in Refs. \cite{gravity,gravity2}. In Ref. \cite{data}, one can find the current limits on the coefficients of the Lorentz symmetry violation.

In this work, we discuss a possible scenario of the Lorentz symmetry violation based on the arising of geometric quantum phases yielded by the effects of the Lorentz symmetry violation in the CPT-even gauge sector of the Standard Model Extension. Geometric quantum phases were termed by Berry \cite{berry} to describe the phase shift acquired by the wave function of a quantum particle in an adiabatic cyclic evolution. At present days, it is well-known that geometric quantum phases can be measured in any cyclic evolution \cite{ahan,anan4,anan5,tg1}. From the effects of the Lorentz symmetry violation in the CPT-even gauge sector of SME, we establish a possible scenario where analogues of the Anandan geometric quantum phase \cite{anan,anan2} and the He-McKellar-Wilkens effect \cite{hmw} can stem from.  Besides, we confine a neutral particle to a hard-wall confining potential and show that the energy levels depend on the He-McKellar-Wilkens geometric phase induced by Lorentz symmetry breaking effects.

The structure of this paper is: in section II, we introduce the non-minimal coupling that describes the Lorentz symmetry breaking in the CPT-even Gauge Sector of the Standard Model Extension and obtain the analogues of the Anandan quantum phase \cite{anan,anan2} and the He-McKellar-Wilkens effect \cite{hmw}; in section III, we build quantum holonomies based on the analogue of the Anandan quantum phase and discuss a possible analogy with the holonomic/geometric quantum computation \cite{zr1,zr2,vedral,vedral2,vedral3,hqc,hqc2,hqc3}; in section IV, we confine a neutral particle to a hard-wall confining potential and discuss the dependence of the energy levels on the He-McKellar-Wilkens geometric phase induced by Lorentz symmetry breaking effects; in section V, we present our conclusions.

\section{analogue of the He-McKellar-Wilkens effect}

In this section, we discuss the effects of the Lorentz symmetry breaking in the CPT-even Gauge Sector of the Standard Model Extension in the nonrelativistic limit of the Dirac equation for a neutral particle. We show that analogues of the Anandan quantum phase \cite{anan,anan2} and the He-McKellar-Wilkens effect \cite{hmw} can be obtained from the effects of the violation of the Lorentz symmetry. Recently, a non-minimal coupling has been introduced in the fermionic sector whose property is that a tensor governs the Lorentz symmetry violation in the CPT-even gauge sector of the Standard Model Extension \cite{manoel,manoel2}. The coupling suggested allow us to write the Dirac equation in the form:
\begin{eqnarray}
m\,\psi=i\gamma^{\mu}\partial_{\mu}\psi+\frac{ig}{2}\,\gamma^{\mu}\,\left(k_{F}\right)_{\mu\nu\alpha\beta}\,\gamma^{\nu}\,F^{\alpha\beta}\left(x\right)\psi,
\label{1.1}
\end{eqnarray}
where $g$ is a constant, $\left(k_{F}\right)_{\mu\nu\alpha\beta}$ is the tensor that governs the Lorentz symmetry violation in the CPT-even gauge sector of the Standard Model Extension. Furthermore, the tensor $\left(k_{F}\right)_{\mu\nu\alpha\beta}$ can be written in terms of four $3\times3$ matrices defined as
\begin{eqnarray}
\left(\kappa_{DE}\right)_{ij}&=&-2\left(k_{F}\right)_{0j0k};\nonumber\\
\left(\kappa_{HB}\right)_{jk}&=&\frac{1}{2}\epsilon_{jpq}\,\epsilon_{klm}\left(k_{f}\right)^{pqlm}\\
\left(\kappa_{DB}\right)_{jk}&=&-\left(\kappa_{HE}\right)_{kj}=\epsilon_{kpq}\left(k_{F}\right)^{0jpq}.\nonumber
\label{1.2}
\end{eqnarray}

Observe that the matrices $\left(\kappa_{DE}\right)_{ij}$ and $\left(\kappa_{HB}\right)_{ij}$ are symmetric and represent the parity-even sector of the tensor $\left(k_{F}\right)_{\mu\nu\alpha\beta}$. On the other hand, the matrices $\left(\kappa_{DB}\right)_{ij}$ and $\left(\kappa_{HE}\right)_{ij}$ has no symmetry and represent the parity-odd sector of the tensor $\left(k_{F}\right)_{\mu\nu\alpha\beta}$. Our focus in this work is on the effects yielded by the parity-odd sector of the tensor $\left(k_{F}\right)_{\mu\nu\alpha\beta}$. In this way, we consider all components $\left(\kappa_{DE}\right)_{ij}$ being zero from now on. The tensor $F_{\mu\nu}\left(x\right)$ in (\ref{1.1}) is the usual electromagnetic tensor ($F_{0i}=-F_{i0}=E_{i}$, and $F_{ij}=-F_{ji}=\epsilon_{ijk}B^{k}$), and the $\gamma^{\mu}$ matrices are defined in the Minkowski spacetime in the form \cite{greiner}:
\begin{eqnarray}
\gamma^{0}=\hat{\beta}=\left(
\begin{array}{cc}
1 & 0 \\
0 & -1 \\
\end{array}\right);\,\,\,\,\,\,
\gamma^{i}=\hat{\beta}\,\hat{\alpha}^{i}=\left(
\begin{array}{cc}
 0 & \sigma^{i} \\
-\sigma^{i} & 0 \\
\end{array}\right);\,\,\,\,\,\,\Sigma^{i}=\left(
\begin{array}{cc}
\sigma^{i} & 0 \\
0 & \sigma^{i} \\	
\end{array}\right),
\label{1.3}
\end{eqnarray}
with $\vec{\Sigma}$ being the spin vector. The matrices $\sigma^{i}$ correspond to the Pauli matrices and satisfy the relation $\left(\sigma^{i}\,\sigma^{j}+\sigma^{j}\,\sigma^{i}\right)=2\eta^{ij}$. 

Now, let us proceed by extending our discussion to curvilinear coordinates. In this case, we need to apply a coordinate transformation $\frac{\partial}{\partial x^{\mu}}=\frac{\partial \bar{x}^{\nu}}{\partial x^{\mu}}\,\frac{\partial}{\partial\bar{x}^{\nu}}$, and a unitary transformation on the wave function $\psi\left(x\right)=U\,\psi'\left(\bar{x}\right)$ \cite{schu,bbs3,bb2}. Thereby, the Dirac equation (\ref{1.1}) can be written in any orthogonal system in the following form:
\begin{eqnarray}
i\,\gamma^{\mu}\,D_{\mu}\,\psi+\frac{i}{2}\,\sum_{k=1}^{3}\,\gamma^{k}\,\left[D_{k}\,\ln\left(\frac{h_{1}\,h_{2}\,h_{3}}{h_{k}}\right)\right]\psi+\frac{ig}{2}\gamma^{\mu}\left(k_{F}\right)_{\mu\nu\alpha\beta}\,\gamma^{\nu}\,F^{\alpha\beta}\left(x\right)\psi=m\psi,
\label{1.4}
\end{eqnarray}
where $D_{\mu}=\frac{1}{h_{\mu}}\,\partial_{\mu}$ is the derivative of the corresponding coordinate system, and the parameters $h_{k}$ correspond to the scale factors of this coordinate system \cite{schu}. As an example, let us consider the Minkowski spacetime, whose line element is written in cylindrical coordinates as: $ds^{2}=-dt^{2}+d\rho^{2}+\rho^{2}d\varphi^{2}+dz^{2}$; then, the corresponding scale factors are $h_{0}=1$, $h_{1}=1$, $h_{2}=\rho$ and $h_{3}=1$. Moreover, the second term in (\ref{1.4}) gives rise to a term called the spinorial connection $\Gamma_{\mu}\left(x\right)$ \cite{schu,bbs2,bbs3}. Hence, the Dirac equation under the influence of the the parity-odd sector of the tensor $\left(k_{F}\right)_{\mu\nu\alpha\beta}$ becomes (we consider $\hbar=c=1$)
\begin{eqnarray}
m\psi=i\gamma^{0}\frac{\partial\psi}{\partial t}+i\gamma^{1}\left(\frac{\partial}{\partial\rho}+\frac{1}{2\rho}\right)\psi+i\frac{\gamma^{2}}{\rho}\,\frac{\partial\psi}{\partial\varphi}+i\gamma^{3}\frac{\partial\psi}{\partial z}+ig\,\vec{\alpha}\cdot\vec{\mathbf{B}}\,\psi-g\,\vec{\Sigma}\cdot\vec{\mathbf{E}}\,\psi,
\label{1.5}
\end{eqnarray}
where the effective fields given in the Dirac equation (\ref{1.5}) are defined as 
\begin{eqnarray}
\mathbf{E}_{i}=\left(\kappa_{HE}\right)_{ij}\,E^{j};\,\,\,\,\,\,\mathbf{B}_{i}=\left(\kappa_{DB}\right)_{ij}\,B^{j}.
\label{1.6}
\end{eqnarray}

Henceforth, we focus on the nonrelativistic limit of the Dirac equation (\ref{1.5}). The nonrelativistic limit of the Dirac equation can be obtained by extracting the temporal dependence of the wave function due to the rest energy \cite{greiner}; thus, we write the Dirac spinor in the form 
\begin{eqnarray}
\psi=e^{-imt}\left(\eta\,\,\,\chi\right)^{T},
\label{1.6a}
\end{eqnarray}
where $\eta$ and $\chi$ are two-component spinors corresponding to the ``large'' and ``small'' components, respectively. Substituting this solution into the Dirac equation (\ref{1.5}), we obtain two coupled equations for $\eta$ and $\chi$, where the first coupled equation is
\begin{eqnarray}
i\frac{\partial\eta}{\partial t}-g\,\vec{\sigma}\cdot\vec{\mathbf{E}}\,\eta=\left[\vec{\sigma}\cdot\vec{\pi}-ig\,\vec{\sigma}\cdot\vec{\mathbf{B}}\right]\chi,
\label{1.7}
\end{eqnarray}
where $\vec{\pi}=\vec{p}-i\vec{\xi}$ and $-i\xi_{k}=-\frac{1}{2\rho}\,\sigma^{3}\,\delta_{2k}$ \cite{bbs3,bb2}. The second coupled equation is
\begin{eqnarray}
2m\chi+i\frac{\partial\chi}{\partial t}+g\,\vec{\sigma}\cdot\vec{\mathbf{E}}\,\chi=\left[\vec{\sigma}\cdot\vec{\pi}+ig\,\vec{\sigma}\cdot\vec{\mathbf{B}}\right]\eta.
\label{1.8}
\end{eqnarray}

Since $\chi$ is the small component of the Dirac spinor, then, we can take $\left|2m\chi\right|\gg\left|i\frac{\partial\chi}{\partial t}\right|$ and  $\left|2m\chi\right|\gg\left|g\,\vec{\sigma}\cdot\vec{\mathbf{E}}\right|$. Thus, we can write
\begin{eqnarray}
\chi\approx\frac{1}{2m}\left[\vec{\sigma}\cdot\vec{\pi}+ig\,\vec{\sigma}\cdot\vec{\mathbf{B}}\right]\eta.
\label{1.9}
\end{eqnarray}
Substituting (\ref{1.9}) into (\ref{1.7}), we obtain
\begin{eqnarray}
i\frac{\partial\eta}{\partial t}=\frac{1}{2m}\left[\vec{p}-i\vec{\xi}+g\left(\vec{\sigma}\times\vec{\mathbf{B}}\right)\right]^{2}\eta-\frac{g^{2}\mathbf{B}^{2}}{2m}\,\eta+\frac{g}{2m}\left(\vec{\nabla}\cdot\vec{\mathbf{B}}\right)\,\eta+g\,\vec{\sigma}\cdot\vec{\mathbf{E}}\,\eta.
\label{1.10}
\end{eqnarray}
which is the Schr\"odinger-Pauli equation based on a Lorentz symmetry breaking scenario defined by the parity-odd sector of the tensor $\left(k_{F}\right)_{\mu\nu\alpha\beta}$. Note that we have analogues of the vector potential and the scalar potential given by
\begin{eqnarray}
\vec{A}_{\mathrm{eff}}=\vec{\sigma}\times\vec{\mathbf{B}};\,\,\,\,\,\,\,\,A^{\mathrm{eff}}_{0}=\vec{\sigma}\cdot\vec{\mathbf{E}}.
\label{1.11}
\end{eqnarray}

Now, we are able to discuss the arising of geometric quantum phases in the wave function of a nonrelativistic Dirac neutral particle yielded by the Lorentz symmetry breaking in the CPT-odd Gauge Sector of the Standard Model Extension. By applying the Dirac phase factor method \cite{dirac,dirac2} to the Schr\"odinger-Pauli equation (\ref{1.10}), where we can write the wave function in the form
\begin{eqnarray}
\eta=\hat{\mathcal{P}}\,e^{i\phi_{\mathrm{A}}}\,\eta_{0},
\label{1.12}
\end{eqnarray}
where  $\hat{\mathcal{P}}$ denotes the path ordering operator and $\eta_{0}$ is the solution to the Schr\"odinger-Pauli equation in the absence of fields, that is, 
\begin{eqnarray}
i\frac{\partial\eta_{0}}{\partial t}=\frac{1}{2m}\left[\vec{p}-i\vec{\xi}\right]^{2}\eta_{0}-\frac{g^{2}\mathbf{B}^{2}}{2m}\,\eta_{0}+\frac{g}{2m}\left(\vec{\nabla}\cdot\vec{\mathbf{B}}\right)\,\eta_{0},
\label{1.13}
\end{eqnarray}
where the term proportional to $\mathbf{B}^{2}$ of Eq. (\ref{1.13}) is a local term, thus, it does not contribute to the geometric phase \cite{anan,anan2}. By comparing with the non-Abelian gauge field $a_{\nu}=\left(-\vec{\sigma}\cdot\vec{B},\,\vec{\sigma}\times\vec{E}\right)$ investigated by Anandan \cite{anan,anan2}, which yields the arising of a geometric phase given by $\Psi\left(x^{\nu}\right)=\hat{\mathcal{P}}\,\exp\left(-i\,\frac{\mu}{\hbar c}\,\oint\,a_{\nu}\,dx^{\nu}\right)\,\Psi_{0}\left(x^{\nu}\right)$ (where $\vec{\mu}=\mu\,\vec{\sigma}$) in the wave function of a neutral particle possessing a permanent magnetic dipole moment, we have that the effective four-vector potential given in Eq. (\ref{1.11}) gives rise to an analogue of the Anandan geometric phase \cite{anan,anan2}:
\begin{eqnarray}
\phi_{\mathrm{A}}=-g\,\oint\left[\vec{\sigma}\times\vec{\mathbf{B}}\right]\cdot d\vec{r}-g\,\int_{0}^{\tau}\,\vec{\sigma}\cdot\vec{\mathbf{E}}\,dt,
\label{1.14}
\end{eqnarray}
where this analogue of the Anandan quantum phase stems from a Lorentz symmetry breaking scenario defined by the parity-odd sector of the tensor $\left(k_{F}\right)_{\mu\nu\alpha\beta}$. Previous studies have investigated the effects of the Lorentz symmetry violation background and the arising of analogues of the Anandan quantum phase \cite{anan,anan2}. In Ref. \cite{bbs3}, it is shown that the Anandan quantum phase induced by a fixed vector field background corresponds to an Abelian phase. In contrast, the analogue of the Anandan quantum phase obtained in Eq. (\ref{1.4}) is a non-Abelian phase. In agreement with Ref. \cite{abb}, this difference between the Abelian nature of the Anandan quantum phase in Ref. \cite{bbs3} and the non-Abelian nature of the Anandan phase given in Eq. (\ref{1.14}) stems from the Lorentz symmetry violation background being defined by a tensor field in Eq. (\ref{1.1}).

Now, let us consider a field configuration defined by a radial magnetic field produced by a uniform linear distribution of magnetic charges on the $z$-axis \cite{hmw}, that is, 
\begin{eqnarray}
\vec{B}=B^{1}\,\hat{\rho}=\frac{\lambda_{m}}{\rho}\,\hat{\rho},
\label{1.14a}
\end{eqnarray}
where $\lambda_{m}$ is a constant associated with a linear density of magnetic charges, $\rho=\sqrt{x^{2}+y^{2}}$ is the radial coordinate, and $\hat{\rho}$ is a unit vector in the radial direction. Several experiments have been made in order to reproduce the field configuration of the He-McKellar-Wilkens effect \cite{tka,tka2,hmw2}. Besides, it is worth mentioning that Dirac monopoles have been observed in synthetic magnetic field \cite{monomag}. In this case, the Anandan quantum phase (\ref{1.4}) becomes
\begin{eqnarray}
\phi_{A1}=-2\pi\lambda_{m}\,g\left(\kappa_{DB}\right)_{11}\,\sigma^{3}+2\pi\lambda_{m}\,g\left(\kappa_{DB}\right)_{31}\,\sigma^{1}.
\label{1.15}
\end{eqnarray} 

As a particular case, let us consider the matrix $\left(\kappa_{DB}\right)_{ij}$ being a diagonal matrix. From this, the Anandan quantum phase (\ref{1.15}) becomes
\begin{eqnarray}
\phi_{\mathrm{HMW}}=-2\pi\lambda_{m}\,g\left(\kappa_{DB}\right)_{11}\,\sigma^{3},
\label{1.16}
\end{eqnarray} 
which corresponds to the analogue of the He-McKellar-Wilkens effect \cite{hmw} based on a Lorentz symmetry breaking scenario defined by the parity-odd sector of the tensor $\left(k_{F}\right)_{\mu\nu\alpha\beta}$. Observe that the geometric phase (\ref{1.16}) does not depend on the velocity of the Dirac neutral particle which consists in a non-dispersive geometric phase as established in Refs. \cite{disp,disp2,disp3}. Moreover, this analogue of the He-McKellar-Wilkens geometric phase is a non-Abelian phase due to the Lorentz symmetry violation background defined by a tensor field given in Eq. (\ref{1.1}), by contrast, the analogue of the He-McKellar-Wilkens geometric phase obtained in Ref. \cite{bbs3} is an Abelian phase due to the Lorentz symmetry violation background being defined by a fixed vector field.

\section{Quantum holonomies}

In this section, let us build quantum holonomies by using the result obtained in Eq. (\ref{1.16}) and discuss a possible analogy with the holonomic/geometric quantum computation \cite{zr1,vedral,vedral2,vedral3,hqc,hqc2,hqc3}. Holonomy or holonomy transformations are unitary transformations which measure the change in the direction of a vector or a spinor when these mathematical quantities are parallel transported either between two different points via different paths or around a closed loop. In short, a holonomy corresponds to a matrix that represents the parallel transport of vectors, spinors, etc. on a closed path and yields the information about the topology or curvature of a given manifold \cite{hol,hol2,hol3}.

From the proposal made by Zanardi and Raseti \cite{zr1} to implement logical gates in a quantum computer though the Berry phase, the interest in studying holonomies in quantum systems has increased in recent years and this branch of quantum computation is known as the holonomic/geometric quantum computation \cite{zr1,zr2,vedral,vedral2,vedral3,hqc,hqc2,hqc3}. The holonomic quantum computation is defined in the eigenspace spanned by the eigenvectors of a family of Hamiltonian operators $\mathcal{F}=\left\{H\left(\lambda\right)=\mathbb{U}\left(\lambda\right)H_{0}\mathbb{U}^{\dag}\left(\lambda\right);\lambda\in\mathcal{M}\right\}$, where $\mathbb{U}\left(\lambda\right)$ is a unitary operator, and $\lambda$ corresponds to the control parameter that can be changed adiabatically along a loop in the control manifold $\mathcal{M}$. The action of the unitary operator $\mathbb{U}\left(\lambda\right)$ on an initial state $\left|\psi_{0}\right\rangle$ brings it to a final state $\left|\psi\right\rangle=\mathbb{U}\left(\lambda\right)\left|\psi_{0}\right\rangle$ giving rise to a quantum gate \cite{loyd}. The general expression of the action of this unitary operator is given by $\left|\psi\right\rangle=\mathbb{U}\left(\lambda\right)\left|\psi_{0}\right\rangle=e^{-i\int_{0}^{t}E\left(t'\right)\,dt'}\,\Gamma_{A}\left(\lambda\right)\left|\psi_{0}\right\rangle$, where the first terms $e^{-i\int_{0}^{t}E\left(t'\right)\,dt'}$ and $\Gamma_{A}\left(\lambda\right)$ correspond to the dynamical phase and the holonomy, respectively. The object $\mathcal{A}=A\left(\lambda\right)\,d\lambda$ is a connection $1$-form called the Mead-Berry connection 1-form \cite{tg1} and the object $A\left(\lambda\right)$ corresponds to the Mead-Berry vector potential, whose components are defined as: $A^{\alpha\beta}=\left\langle \psi^{\alpha}\left(\lambda\right)\right|\partial/\partial\lambda\left|\psi^{\beta}\left(\lambda\right)\right\rangle$. However, based on Ref. \cite{ahan}, the dynamical phase can be omitted by redefining the energy levels (for instance, by taking $\mathcal{E}\left(0\right)=0$), then, one can study the appearance of geometric phases in any cyclic evolution of the quantum system.

In this work, we have seen that we can obtain the geometric quantum phase (\ref{1.15}) without applying the adiabatic approximation, which agrees with Ref. \cite{ahan}. Besides, by assuming the possibility of detecting Lorentz symmetry breaking effects, then, we can make an analogy with the geometric quantum computation \cite{vedral3,hqc,hqc2,hqc3}. This analogy is based on measuring the uncertainties of the Lorentz symmetry breaking parameters through geometric phases as in Eq. (\ref{1.7}), therefore, we define the control parameters through the terms related to the Lorentz symmetry violation not in the sense that they can be slowly changing parameters, but in the sense that we can know or determine them previously \cite{abb,bb}. Hence, let us define the logical states of this system as being the spin of the Dirac neutral particle, that is,
\begin{eqnarray}
\left|0_{\mathrm{L}}\right\rangle=\left|\uparrow\right\rangle;\,\,\,\,\,\,\,\left|1_{\mathrm{L}}\right\rangle=\left|\downarrow\right\rangle,
\label{2.1}
\end{eqnarray}
where $\left|\uparrow\right\rangle$ and $\left|\downarrow\right\rangle$ correspond to the spin up and the spin down of the Dirac neutral particle (the spin of the neutral particle being initially polarized along the $z$-axis), respectively. Observe that the choice of the logical basis above is justified due to the coupling of the spin of the Dirac neutral particle with the Lorentz symmetry violation background, which is manifested in the geometric phase (\ref{1.15}). Thereby, the parallel transport of spinors given in Eq. (\ref{2.1}) on a closed path is yielded by the quantum holonomy associated with the geometric phase (\ref{1.15}) in a cyclic evolution, which is defined in the form:
\begin{eqnarray}
\mathbb{U}\left(\zeta_{1},\zeta_{3}\right)=\exp\left(-i\zeta_{3}\,\sigma^{3}+i\zeta_{1}\,\sigma^{1}\right),
\label{2.2}
\end{eqnarray}
where we have defined the parameters
\begin{eqnarray}
\zeta_{1}=2\pi\lambda_{m}\,g\left(\kappa_{DB}\right)_{31};\,\,\,\,\,\zeta_{3}=2\pi\lambda_{m}\,g\left(\kappa_{DB}\right)_{11}.
\label{2.3}
\end{eqnarray}

The holonomy transformation (\ref{2.2}) has the sum of two non-commuting matrices into the argument of the exponential function, thus, we have that $e^{A+B}\neq e^{A}\,e^{B}$. Thus, in order to simplify the expression of the unitary operator (\ref{2.2}) acting on the logical states (\ref{2.1}), we use the relation: $e^{A+B}=e^{A}\,e^{B}\,e^{-\frac{1}{2}\left[A,B\right]}\cdots$ (where $A$ and $B$ are matrices). In this way, we obtain:  
\begin{eqnarray}
\mathbb{U}\left(\zeta_{1},\zeta_{3}\right)\approx e^{-i\zeta_{3}\,\sigma^{3}}\,e^{i\zeta_{1}\,\sigma^{1}}\,e^{-i\zeta_{3}\zeta_{1}\sigma^{2}},
\label{2.4}
\end{eqnarray}
where we have neglected terms of order $\mathcal{O}\left(\zeta_{1}^{2}\,\zeta_{3}\right)$, $\mathcal{O}\left(\zeta_{3}^{2}\,\zeta_{1}\right)$ and higher, because we can consider these terms very small. Moreover, by using the definition of the function of a matrix, that is, $\exp{A}=\sum_{i=0}^{\infty}\frac{A^{n}}{n!}$, we can write (\ref{2.4}) as
\begin{eqnarray}
\mathbb{U}\left(\zeta_{1},\zeta_{3}\right)\approx \omega_{0}\,I+i\,\omega_{1}\,\sigma^{1}-i\,\omega_{2}\,\sigma^{2}+i\,\omega_{3}\,\sigma^{3},
\label{2.5}
\end{eqnarray} 
where the parameters $\omega_{k}$ given in (\ref{2.5}) are
\begin{eqnarray}
\omega_{0}&=&\cos\zeta_{3}\,\cos\zeta_{1}\,\cos\zeta_{1}\zeta_{3}+\sin\zeta_{1}\,\sin\zeta_{3}\,\sin\zeta_{1}\zeta_{3};\nonumber\\
\omega_{1}&=&\cos\zeta_{3}\,\sin\zeta_{1}\,\cos\zeta_{1}\zeta_{3}+\sin\zeta_{3}\,\cos\zeta_{1}\,\sin\zeta_{1}\zeta_{3};\nonumber\\
[-3mm]\label{2.6}\\[-3mm]
\omega_{2}&=&\cos\zeta_{3}\,\cos\zeta_{1}\,\sin\zeta_{1}\zeta_{3}-\sin\zeta_{3}\,\sin\zeta_{1}\,\cos\zeta_{1}\zeta_{3};\nonumber\\
\omega_{3}&=&\cos\zeta_{3}\,\sin\zeta_{1}\,\sin\zeta_{1}\zeta_{3}-\sin\zeta_{3}\,\cos\zeta_{1}\,\cos\zeta_{1}\zeta_{3}.\nonumber
\end{eqnarray}

Finally, we can complete the analogy with geometric/holonomic quantum computation \cite{zr1,zr2,vedral,vedral2,vedral3,hqc,hqc2,hqc3} by considering the parameters associated with the Lorentz symmetry violation $\zeta_{3}$ and $\zeta_{1}$ given in Eq. (\ref{2.3}) as control parameters. These parameters can be considered as control parameters in the sense that we can know or determine the values of $g\left(\kappa_{DB}\right)_{11}$ and $g\left(\kappa_{DB}\right)_{31}$ previously, based on measuring the uncertainty of the Lorentz symmetry violation parameters via geometric quantum phases, which allow us to estimate the values of the Lorentz symmetry violation parameters, and by choosing the appropriate values of $\lambda_{m}$ in an analogous way to an electric charge density \cite{bf11}. From this perspective, we can apply the unitary transformation defined by Eq. (\ref{2.5}) to the logical states (\ref{2.1}), which means that we can make a rotation on the logical states (\ref{2.1}) through the appropriate choice of the control parameters $\zeta_{1}$ and $\zeta_{3}$ given in Eq. (\ref{2.3}).  By applying the holonomy transformation (\ref{2.5}) on the logical states (\ref{2.1}) several times, then, we can perform a universal set of one-qubit quantum gates and implement the geometric/holonomic quantum computation \cite{vedral,vedral2,vedral3,hqc,hqc2,hqc3,zr1} based on the Lorentz symmetry breaking in the CPT-odd Gauge Sector of the Standard Model Extension.

\section{confinement to a hard-wall confining potential}

In this section, let us consider the nonrelativistic Dirac particle under the influence of the Lorentz symmetry breaking effects that give rise to the geometric phase (\ref{1.16}). Then, let us discuss a case where this quantum particle is confined to a hard-wall confining potential and investigate the quantum effects associated with the analogue of the He-McKellar-Wilkens effect \cite{hmw}. Observe that, from Eqs. (\ref{1.11}) and (\ref{1.16}), we can write
\begin{eqnarray}
g\,\vec{A}_{\mathrm{eff}}=g\,\vec{\sigma}\times\mathbf{B}=\frac{\phi_{\mathrm{HMW}}}{\phi_{0}\,\rho}\,\sigma^{3}\,\hat{\varphi},
\label{4.1}
\end{eqnarray}
where $\hat{\varphi}$ is a unit vector in the azimuthal direction and $\phi_{0}=2\pi$. In this way, the Schr\"odinger-Pauli equation (\ref{1.10}) becomes
\begin{eqnarray}
i\frac{\partial\eta}{\partial t}&=&-\frac{1}{2m}\left[\frac{\partial^{2}}{\partial\rho^{2}}+\frac{1}{\rho}\,\frac{\partial}{\partial\rho}+\frac{1}{\rho^{2}}\,\frac{\partial^{2}}{\partial\varphi^{2}}+\frac{\partial^{2}}{\partial z}\right]\eta+\frac{1}{2m}\,\frac{i\sigma^{3}}{\rho^{2}}\,\frac{\partial\eta}{\partial\varphi}+\frac{1}{8m\rho^{2}}\,\eta\nonumber\\
[-2mm]\label{4.2}\\[-2mm]
&-&\frac{i}{m}\,\frac{\phi_{\mathrm{HMW}}}{\phi_{0}\,\rho^{2}}\,\frac{\partial\eta}{\partial\varphi}+\frac{1}{2m}\,\left(\frac{\phi_{\mathrm{HMW}}}{\phi_{0}\rho}\right)^{2}\,\eta-\frac{1}{2m}\,\frac{\phi_{\mathrm{HMW}}}{\phi_{0}\rho^{2}}\,\eta.\nonumber
\end{eqnarray}

Observe that the operators $\hat{p}_{z}=-i\partial_{z}$ and $\hat{J}_{z}=-i\partial_{\varphi}$ \cite{schu} commute with the Hamiltonian of the right-hand side of Eq. (\ref{4.2}). Moreover, we can also see in Eq. (\ref{4.2}) that $\eta$ is an eigenfunction of $\sigma^{3}$, whose eigenvalues are $s=\pm1$. Thereby, we write $\sigma^{3}\eta_{s}=\pm\eta_{s}=s\eta_{s}$. Hence, the solution to Eq. (\ref{4.2}) can be written in terms of the eigenvalues of the operators above: 
\begin{eqnarray}
\eta_{s}=e^{-i\mathcal{E}t}\,e^{i\left(l+\frac{1}{2}\right)\varphi}\,e^{ikz}\,R_{s}\left(\rho\right),
\label{4.3}
\end{eqnarray}
where $l=0,\pm1,\pm2,\ldots$ and $k$ is a constant. Substituting the solution (\ref{4.3}) into the Schr\"odinger-Pauli equation (\ref{4.2}), we obtain
\begin{eqnarray}
R_{s}''+\frac{1}{\rho}R_{s}'-\frac{\tau^{2}}{\rho^{2}}R_{s}+\beta^{2}\,R_{s}=0,
\label{4.4}
\end{eqnarray}
where we have defined the following parameters in Eq. (\ref{4.4}):
\begin{eqnarray}
\tau&=&l+\frac{1}{2}\left(1-s\right)+s\,\frac{\phi_{\mathrm{HMW}}}{\phi_{0}};\nonumber\\
[-2mm]\label{4.5}\\[-2mm]
\beta^{2}&=&2m\mathcal{E}-k^{2}.\nonumber
\end{eqnarray} 

From now on, let us take $k=0$ in order to have a planar system. Note that the radial equation (\ref{4.3}) corresponds to the Bessel differential equation \cite{abra}, whose general solution is $R_{s}\left(\rho\right)=A\,J_{\tau}\left(\beta\rho\right)+B\,N_{\tau}\left(\beta\rho\right)$, where the functions $J_{\tau}\left(\beta\rho\right)$ and $N_{\tau}\left(\beta\rho\right)$ are the Bessel functions of the first and second kinds. 

Henceforth, we consider the wave function of the particle is well-behaved at the origin and vanishes at a fixed radius $\rho_{0}$. Since the function $N_{\tau}\left(\beta\rho\right)$ diverges at the origin, therefore, we must take $B=0$ and write the solution to Eq. (\ref{4.3}) as $R_{s}\left(\rho\right)=A\,J_{\left|\tau\right|}\left(\beta\rho\right)$ \cite{moraes}. Thereby, by assuming $\beta\rho_{0}\gg0$, then, we can write \cite{abra,moraes}
\begin{eqnarray}
J_{\left|\tau\right|}\left(\beta\rho_{0}\right)\rightarrow\sqrt{\frac{2}{\pi\beta\rho_{0}}}\,\cos\left(\beta\rho_{0}-\frac{\left|\tau\right|\,\pi}{2}-\frac{\pi}{4}\right).
\label{4.6}
\end{eqnarray}

Therefore, from Eqs. (\ref{4.5}) and (\ref{4.6}) and by imposing $R_{s}\left(\rho_{0}\right)=0$, we obtain
\begin{eqnarray}
\mathcal{E}_{n,\,l,\,s}\approx\frac{1}{2m\rho_{0}^{2}}\,\left[n\pi+\frac{\pi}{2}\,\left|l+\frac{1}{2}\left(1-s\right)+s\,\frac{\phi_{\mathrm{HMW}}}{\phi_{0}}\right|+\frac{3\pi}{4}\right]^{2},
\label{4.7}
\end{eqnarray}
where $n=0,1,2,\ldots$ is the quantum number associated with the radial modes. 

Hence, the spectrum of energy given in Eq. (\ref{4.7}) is the energy levels of a spin-$1/2$ neutral particle under the influence of the Lorentz symmetry breaking in the CPT-odd Gauge Sector of the Standard Model Extension confined to a hard-wall confining potential. Note that the energy levels (\ref{4.7}) depend on the analogue of the He-McKellar-Wilkens geometric phase given in Eq. (\ref{1.16}) whose periodicity is $\phi_{0}=2\pi$, that is, we have that $\mathcal{E}_{n,\,l,\,s}\left(\phi_{\mathrm{HMW}}+\phi_{0}\right)=\mathcal{E}_{n,\,l+1,\,s}\left(\phi_{\mathrm{HMW}}\right)$.

\section{conclusions}

We have shown in this work that analogues of the He-McKellar-Wilkens effect \cite{hmw} and the Anandan quantum phase for a neutral particle with a permanent electric dipole moment \cite{anan,anan2} can be yielded by Lorentz symmetry breaking effects in the CPT-even gauge sector of the Standard Model Extension. These possible scenarios of the violation of the Lorentz symmetry have allowed us to obtain a phase shift in the wave function of the neutral particle which does not depend on the velocity of the particle, a non-dispersive phase \cite{disp,disp2,disp3}, and has a non-Abelian nature. This non-Abelian nature of the geometric phases stems from the Lorentz symmetry breaking scenario defined by the parity-odd sector of the tensor $\left(k_{F}\right)_{\mu\nu\alpha\beta}$. 

From the analogue of the Anandan quantum phase \cite{anan,anan2}, we have build quantum holonomies in which a possible analogy with the geometric/holonomic quantum computation can be make. This analogy is based on the assumption that Lorentz symmetry breaking effects can be detected from the measure of the uncertainties of the Lorentz symmetry breaking parameters through a geometric phase. Therefore, we can define the control parameters through the terms related to the Lorentz symmetry violation in the sense that we can know or determine them previously, but not in the sense that they can be slowly changing parameters \cite{abb,bb}.

Finally, we have investigated the Lorentz symmetry breaking effects on the neutral particle confined to a hard-wall confining potential. We have chosen the scenario of the Lorentz symmetry violation that gives rise to the arising of a geometric phase analogous to the He-McKellar-Wilkens geometric phase \cite{hmw} and shown that the energy levels depend on this geometric quantum phase.

\acknowledgments{The authors would like to thank CNPq (Conselho Nacional de Desenvolvimento Cient\'ifico e Tecnol\'ogico - Brazil) for financial support.}

\end{document}